\documentclass[aps,prl,reprint,superscriptaddress,notitlepage,longbibliography]{revtex4-2}

\usepackage{amssymb,amsmath}
\usepackage{graphicx}
\usepackage{dcolumn}
\usepackage{multirow}
\usepackage[english]{babel}
\usepackage{bm}
\usepackage[utf8]{inputenc}
\usepackage[T1]{fontenc}
\usepackage{mathptmx}
\usepackage{etoolbox}
\usepackage{braket}
\usepackage[colorlinks=true, linkcolor=black, citecolor=blue, urlcolor=blue]{hyperref}
\usepackage{xr}
\usepackage{color}
\usepackage{color,soul} 
\sethlcolor{white}
\soulregister\cite7
\soulregister\ref7
\usepackage{textcomp}
\usepackage{anyfontsize}

\begin{document}

\title{\textbf{\fontfamily{phv}\selectfont 
Large quantum dot energy level shifts in anomalous photon-assisted tunneling}}
\author{Jared Benson}
\email{jcbenson2@wisc.edu}
\affiliation{Department of Physics, University of Wisconsin-Madison, Madison, WI 53706, USA}

\author{C. E. Sturner}
\affiliation{Department of Physics, University of Wisconsin-Madison, Madison, WI 53706, USA}

\author{A. R. Huffman}
\affiliation{Department of Physics, University of Wisconsin-Madison, Madison, WI 53706, USA}

\author{Sanghyeok Park}
\affiliation{Department of Physics, University of Wisconsin-Madison, Madison, WI 53706, USA}

\author{Valentin John}
\affiliation{QuTech and Kavli Institute of Nanoscience, Delft University of Technology, P.O. Box 5046, 2600 GA Delft, The Netherlands}

\author{Brighton X. Coe}
\affiliation{Department of Physics, University of Wisconsin-Madison, Madison, WI 53706, USA}

\author{Tyler J. Kovach}
\affiliation{Department of Physics, University of Wisconsin-Madison, Madison, WI 53706, USA}

\author{Stefan D. Oosterhout}
\affiliation{QuTech and Netherlands Organisation for Applied Scientific Research, 2628 CK Delft, The Netherlands}

\author{Lucas E. A. Stehouwer}
\affiliation{QuTech and Kavli Institute of Nanoscience, Delft University of Technology, P.O. Box 5046, 2600 GA Delft, The Netherlands}

\author{Francesco Borsoi}
\thanks{Present address: Niels Bohr Institute,
University of Copenhagen, Blegdamsvej 17, 2100 Copenhagen, Denmark.}
\affiliation{QuTech and Kavli Institute of Nanoscience, Delft University of Technology, P.O. Box 5046, 2600 GA Delft, The Netherlands}

\author{Giordano Scappucci}
\affiliation{QuTech and Kavli Institute of Nanoscience, Delft University of Technology, P.O. Box 5046, 2600 GA Delft, The Netherlands}

\author{Menno Veldhorst}
\affiliation{QuTech and Kavli Institute of Nanoscience, Delft University of Technology, P.O. Box 5046, 2600 GA Delft, The Netherlands}

\author{Benjamin D. Woods}
\affiliation{Department of Physics, University of Wisconsin-Madison, Madison, WI 53706, USA}

\author{Mark Friesen}
\affiliation{Department of Physics, University of Wisconsin-Madison, Madison, WI 53706, USA}

\author{M. A. Eriksson}
\affiliation{Department of Physics, University of Wisconsin-Madison, Madison, WI 53706, USA}

\begin{abstract}
Orbital energy splittings are important quantum dot parameters for the operation of hole spin qubits. They are known to depend on the lateral confinement of the quantum dots. However, when changing top, plunger gate voltages, which are the typical control parameter for qubit applications, such energy splitting changes are typically negligible, both as measured in experiment and as assumed in effective theories. Here, we study the singlet-triplet (ST) splittings, which depend on the orbital splittings, of a double quantum dot (DQD) in a Ge/SiGe heterostructure using photon-assisted tunneling (PAT) and pulsed-gate spectroscopy. We find that the ST splittings have a surprising, strong dependence on the top gate voltages, leading to anomalous PAT measurements. We combine data from both measurements in a model that well describes the linear gate-voltage dependence of the ST splittings. Finally, we show that the ST splittings of the two dots exhibit similar linear gate-voltage dependences when the device is retuned such that their ratio is significantly different.
\end{abstract}
\maketitle

The orbital energy splittings of holes in quantum dots formed in Ge/SiGe heterostructures set important energy scales for the operation of hole spin qubits. For systems comprised of multiple quantum dots, including multiple single-hole spin qubits~\cite{Hendrickx2020-bs, Wang2022-gf} or multi-hole spin qubits such as singlet-triplet qubits~\cite{Jirovec2021-qa, Zhang2025-km,Tsoukalas2026-wb}, the orbital splittings dictate the singlet-triplet (ST) splittings. These ST splittings determine relevant quantum dot and qubit properties, such as the size of the qubit readout window when performing spin-to-charge conversion~\cite{Hendrickx2020-bs,Jirovec2021-qa}. For the quantum dot hybrid qubit, which has previously been demonstrated with electrons in Si/SiGe~\cite{Kim2014-gb,Kim2015-np} and GaAs~\cite{Cao2016-cp,Jang2021-rv} but not with holes in Ge/SiGe, this ST splitting is even more important, as it determines the qubit energy splitting. Given their importance for qubit manipulation and readout, it is crucial that the ST splittings are tunable in a range that can give rise to easy qubit manipulation and large qubit readout windows.

For electrons in Si/SiGe, ST splittings can be uncontrollable and sometimes vanishingly small due to low-lying valley states~\cite{Zwanenburg2013-iu}. Holes in Ge/SiGe do not have valley states, so hole spin qubits in this platform avoid this difficulty. Instead, ST splittings in Ge/SiGe are typically amply large for qubit operations~\cite{Watzinger2018-aj,Hendrickx2020-pd,Hendrickx2020-bs,Sommer2026-rn}. It is well known that orbital energy splittings, and therefore ST splittings, are determined by the electrostatic environment of the quantum dots; however, in the vicinity of triple points on quantum dot charge stability diagrams, orbital energy levels are observed in experiment to be constant~\cite{van-der-Wiel2002-uo,Hanson2007-ox,Zwanenburg2013-iu}. Furthermore, for this reason most effective qubit theories assume constant values for these parameters~\cite{Burkard2023-zc}.

Here, we study the ST splittings of a double quantum dot (DQD) in a Ge/SiGe heterostructure~\cite{Lodari2021-sb}, finding a surprising, strong variation in ST splittings as a function of applied plunger gate voltages. We use photon-assisted tunneling (PAT)~\cite{Kouwenhoven1994-uj,Blick1995-nk,Fujisawa1997-gz} to measure the three-hole double-dot energy dispersion near the (2,1) - (1,2) charge anticrossing, which depends on both the ST splitting of each dot and the coupling between the dots. As a complementary measurement, we use pulsed-gate spectroscopy~\cite{Elzerman2004-ky,Xiao2010-kw,Yang2012-aa,Zajac2016-od,Liles2018-cr,Gachter2022-av} to measure the ST splitting of each dot away from the (2,1) - (1,2) charge anticrossing. We find that both PAT and pulsed-gate spectroscopy data show that the ST splittings shift by a large amount when making small changes to the plunger gate voltages. To consistently analyze data from both measurements, we use a modified model for the double-dot energy levels in which the ST splittings in each dot vary linearly with gate voltages. We find that this model has excellent agreement with the experimental data. We study this phenomenon across multiple, substantially different device tunings, and in all cases we find similarly large, linear shifts in the ST splittings. Thus, these results are directly relevant to qubit manipulation and readout for hole spin qubits in Ge/SiGe heterostructures that utilize the ST splitting.

\begin{figure}[ht!]
\includegraphics[width=3.375in]{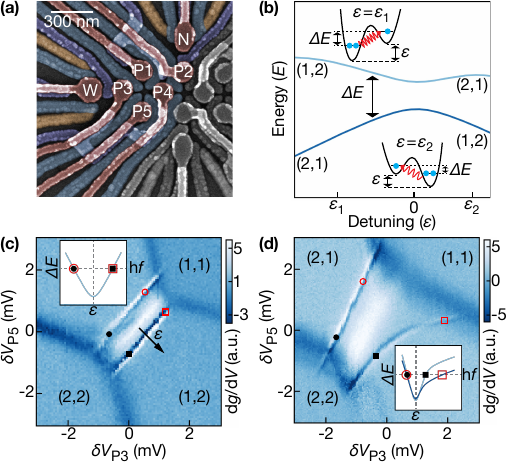}
\caption{\label{fig:fig1} Device layout and anomalous photon-assisted tunneling (PAT). (a) A false-color scanning electron microscope (SEM) image of a device that is nominally identical to the one measured here. Adapted from the raw SEM image in Ref.~\cite{John2025-ml}. (b) An energy dispersion diagram near the (2,1) - (1,2) charge anticrossing. PAT resonances appear when the energy difference between the states ($\Delta E$) is equal to the microwave photon energy (insets). (c) A charge stability diagram measured in the presence of a 15 GHz microwave excitation. PAT resonances are straight because the device is tuned such that the excited state anticrossings occur far away in detuning ($\varepsilon$).  (inset) A diagram of $\Delta E$ between the two black (red) points on the stability diagram. $\Delta E$ is linear on either side of the minimum due to the device tuning described above. (d) The same as (c), except the microwave excitation has a frequency of 18.5 GHz and the device is tuned such that the excited state anticrossings occur closer to zero detuning. The right dot ST splitting is tuned to be small, and it varies with plunger gate voltages, so the anomalous PAT resonance line appearing in (1,2) is curved. The curves in the inset differ from each other and have nonzero slopes at larger absolute detuning ($\vert\varepsilon\vert$) due to the gate-voltage dependence of the ST splittings.}
\end{figure}

We form the DQD in a two-dimensional quantum dot array device~\cite{Wang2024-vg}. Figure~\ref{fig:fig1}a is a false-color scanning electron microscope (SEM) image of a device with the same gate structure as the one measured in this work. The false-color SEM image shown here is adapted from the raw, uncropped and uncolored SEM image in Ref.~\cite{John2025-ml}. We operate only a subset of this device and relevant gates are colored as follows: plunger gates (red), barrier gates (light blue), screening gates (dark blue), and ohmic contacts (yellow). We tune up the DQD under P3 (left dot) and P5 (right dot), which we measure with a charge sensor under W. Holes tunnel into the double dot from the charge sensor under W or from N via P2 and P4.

We tune the DQD to the (2,1) - (1,2) charge configuration and measure the energy difference between these charge states using PAT. Figure~\ref{fig:fig1}b is an energy dispersion diagram of the (2,1) and (1,2) charge states near their anticrossing. The insets illustrate the PAT process: when the energy difference between the (2,1) and (1,2) charge states ($\Delta E$), equals the photon energy ($hf$) of an applied microwave excitation, then the microwave photons drive tunneling of one of the holes from one quantum dot to the other. This movement of charge can be measured directly with the charge sensor.

Figure~\ref{fig:fig1}c shows a charge stability diagram of the DQD in the presence of a 15 GHz microwave excitation. PAT resonance lines are visible in the regions of the stability diagram corresponding to the (2,1) and (1,2) charge configurations. In this example, the device is tuned such that the excited state anticrossings in the energy dispersion are far from the ground state anticrossing (see supplemental material section S1). Because of this, the resonance lines are straight and parallel with the polarization line. This behavior — PAT signal parallel to the polarization line — is the commonly observed signature of PAT in DQDs~\cite{Fujisawa1997-gz,Fujisawa1997-jl,van-der-Wiel2002-uo}. The inset shows a diagram of $\Delta E$ along lines between the black (red) points on the stability diagram. Since the resonance lines are straight, $\Delta E$ is identical along each line. Moreover, since the excited state anticrossings occur at larger positive and negative detunings, $\Delta E$ is linear on either side of the minimum.

If we tune the device so the excited state anticrossings are closer to the ground state anticrossing, we observe anomalous features in the PAT resonances, as shown in Fig.~\ref{fig:fig1}d. The resonance line in (1,2) is curved and not parallel with the polarization line. This is unlike the data shown in Fig.~\ref{fig:fig1}c, as well as previous experiments in GaAs~\cite{Petta2004-bc,Schreiber2011-kl,Wang2016-hf}, bilayer graphene~\cite{Hecker2023-ld,Ruckriegel2025-ic}, and carbon nanotubes~\cite{Mavalankar2016-dq}, in which these anomalous features are notably absent. The curvature and slope of the PAT resonance in (1,2) indicates that the energy dispersion varies with both the detuning and the position along the polarization line, which is caused by the strong dependence of the right-dot ST splitting on plunger gate voltages. The inset shows a diagram of $\Delta E$ along lines between the black (red) points on the stability diagram. Since the energy levels shift between these lines, the curves do not overlap as in Fig.~\ref{fig:fig1}c.

\begin{figure*}[ht!]
\centering
\includegraphics[width=5.31in]{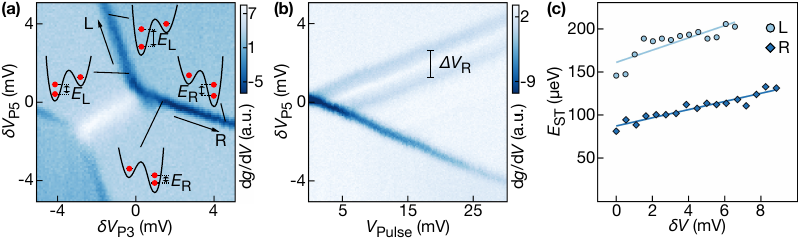}
\caption{\label{fig:fig2} Measuring the dependence of ST splittings on gate voltages with pulsed-gate spectroscopy. (a) A charge stability diagram with insets showing the dependence of the two-hole ST splittings on the plunger gate voltages. Note that the ST splittings increase while moving away from the polarization line. (b) Pulsed-gate spectroscopy measurement to obtain energy splittings. Here, a square wave pulse is applied to the plunger gate P5 while sweeping across a P5 transition line. When the pulse amplitude is sufficiently large, a second loading line appears, and the gap between the two loading lines ($\Delta V_\mathrm{R}$) provides a measurement of the ST splitting in the right dot. (c) Measured ST splittings for the left and right dots as a function of the plunger gate voltages. $\delta V$ increases for the left (right) dot as the measurements move farther from the (2,1) - (1,2) - (1,1) triple point in the direction of the arrow labeled L (R).}
\end{figure*}

We complement the PAT measurements with pulsed-gate spectroscopy, which directly measures the two-hole ST splitting in each dot. Moreover, pulsed-gate spectroscopy is able to measure the ST splitting farther from the (2,1) - (1,2) charge anticrossing (at larger $\vert\varepsilon\vert$), where PAT measurements fail because the ground state and first excited state no longer differ in their charge configurations. Figure~\ref{fig:fig2}b shows an example of a pulsed-gate spectroscopy measurement for the right-dot ST splitting. In this measurement technique, we scan across the reservoir transition line in the stability diagram corresponding to the addition of a second hole to either the left or right dot while applying a square wave pulse. The pulse splits the transition line into a loading line and an unloading line for the ground state. If the pulse amplitude is made sufficiently large, then the excited triplet state becomes accessible and an additional loading line corresponding to this state appears between the ground state loading and unloading lines. By measuring the distance between the two loading lines, we determine the ST splitting.

Figure~\ref{fig:fig2}a is a charge stability diagram with insets that show the trends we observe in the two-hole ST splittings. Moving away from the (2,1) - (1,2) - (1,1) triple point along either transition line results in an increase in the energy splitting. This can be seen in the ST splitting measurements plotted for both dots in Fig.~\ref{fig:fig2}c, where the pulsed-gate spectroscopy measurements are taken along the directions indicated by the arrows in Fig.~\ref{fig:fig2}a. The left-dot ST splitting increases from 146 $\mathrm{\mu eV}$ to 206 $\mathrm{\mu eV}$ and the right-dot ST splitting increases from 81 $\mathrm{\mu eV}$ to 133 $\mathrm{\mu eV}$. The trends in both energy splittings are well described by linear fits (lines in Fig.~\ref{fig:fig2}c).

\begin{figure*}[ht!]
\centering
\includegraphics[width=5.13in]{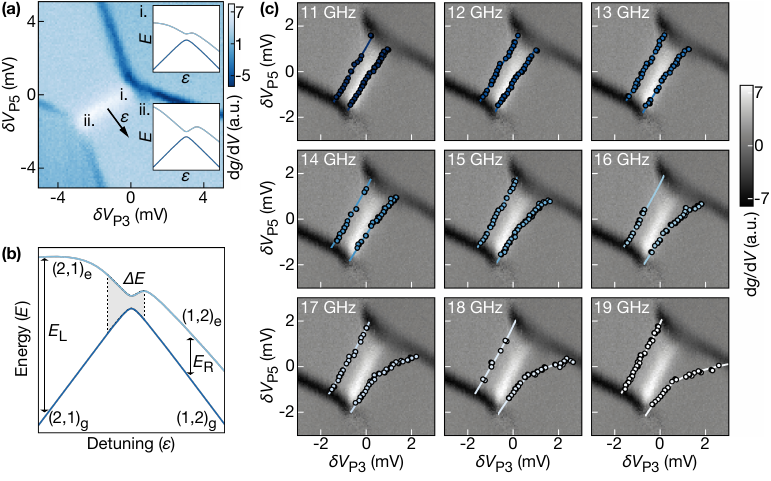}
\caption{\label{fig:fig3} Modeling the energy dispersion using PAT and pulsed-gate spectroscopy data. (a) A charge stability diagram with insets showing the lowest two energy levels of the DQD along lines that are parallel to the detuning axis and intersect the points i. and ii. calculated using the results of the fitting shown in (c). These diagrams show how the energy levels shift due to the gate-voltage dependence of the ST splittings. (b) An energy level diagram showing how the gate-voltage dependence is incorporated into the model used to fit the data. The energy dispersion near zero detuning (highlighted in gray) is fit to the PAT results, which depends on both $V_{\mathrm{P3}}$ and $V_{\mathrm{P5}}$. The energy splittings at larger detunings are set using results from the pulsed-gate spectroscopy measurements shown in Fig.~\ref{fig:fig2}c, and they depend linearly on plunger gate voltages. The left-dot ST splitting ($E_\mathrm{L}$) depends on $V_{\mathrm{P3}}$ and the right dot ST splitting ($E_\mathrm{R}$) depends on both $V_{\mathrm{P5}}$ and $V_{\mathrm{P3}}$ because it is small enough that its gate-voltage dependence is visible in the PAT measurements. An important consequence of the gate-voltage dependence of the ST splittings is that the energy levels never become parallel at large positive or negative detunings. (c) Charge stability diagrams overlaid with experimental PAT data points for specified microwave excitation frequencies. In each subplot, the lines are the contours of best fit from the global fit to the PAT data as described in the main text.}
\end{figure*}

We connect the PAT and pulsed-gate spectroscopy measurements using a model that takes into account the gate-voltage dependence of the ST splittings, as shown in Fig.~\ref{fig:fig3}. Figure~\ref{fig:fig3}a is a charge stability diagram with insets showing the energy dispersions passing through the points i. and ii. The energy levels are calculated using the model described in section S4 of the supplemental material, and the parameter values used in these calculations come from the PAT data fitting described below. Moving in Fig.~\ref{fig:fig3}a from i. to ii. along the polarization line, the energy splitting between the ground state and the first excited state increases due to the gate-voltage dependence of the ST splittings.

Figure~\ref{fig:fig3}b shows schematically how the gate-voltage dependence is included and how the data from each type of measurement contribute to the model of the DQD energy levels. At small detuning values, the PAT data provide a direct measurement of the energy difference between the (2,1) and (1,2) charge states ($\Delta E$) as a function of both gate voltages. At larger detunings, the energy splittings are determined by the ST splittings measured by pulsed-gate spectroscopy. The model assumes that each ST splitting depends linearly on the plunger gate voltages and is primarily influenced by the plunger gate closest to the quantum dot in question. Small crosstalk effects from a plunger gate to the ST splitting of the opposite dot are also included when the ST splitting is small enough that anomalous features can be seen in the PAT data, as in Fig.~\ref{fig:fig1}d. 

The PAT dataset used in the fitting is comprised of multiple (in this case 20) scans of detuning versus microwave excitation frequency. These measurements are offset from each other such that they span the entire (2,1) - (1,2) polarization line. The PAT resonance frequencies are extracted from each of these individual measurements and combined into a two-dimensional map. When fitting to the model described above, the entire dataset is fit simultaneously, and the model requires only a small number of parameters to successfully perform this global fit. More details about the model and the fitting procedure can be found in section S4 of the supplemental material.

Figure~\ref{fig:fig3}c compares the fit from the model to the experimental measurements of the energy dispersion. Each subplot shows a charge stability diagram overlaid with experimental PAT data points acquired with microwave frequencies near the specified value. The lines are contours of constant energy difference from the fit to the model described above, where the energy difference is equal to the photon energy of the microwave excitation. Comparing the contours to the data points, we find excellent agreement.

\begin{figure}[ht!]
\includegraphics[width=3.375in]{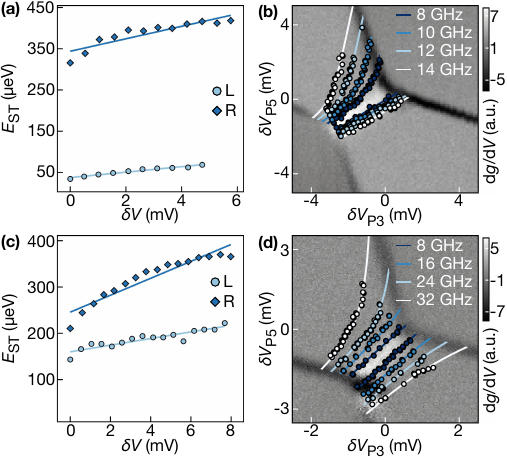}
\caption{\label{fig:fig4} Energy dispersion modeling under different device tunings. (a) ST splittings measured with pulsed-gate spectroscopy in the case that the device is tuned such that the left-dot ST splitting is much smaller than the right-dot ST splitting. (b) PAT resonances compared with fitting results for the tuning described in (a). Since the left-dot ST splitting is small, the PAT resonances are curved in the region of the stability diagram corresponding to the (2,1) charge configuration. (c,d) The same as (a,b), except in a tuning where the ST splittings are more similar in value. Here, the ST splittings are small enough that curvature is present in the PAT resonances in the regions of the stability diagram corresponding to both the (2,1) and (1,2) charge configurations.}
\end{figure}

To ensure that the gate-voltage dependence of the ST splittings is not a consequence of one particular device tuning, we retune the DQD to two additional operating regimes using nearby plunger and barrier gates. Here, we significantly change not only the values of each ST splitting but also the ratio of the two energy splittings. We repeat the same analysis as before, and the results for these additional cases are shown in Fig.~\ref{fig:fig4}.

Figure~\ref{fig:fig4}a,b show results from an opposite device tuning compared to Fig.~\ref{fig:fig3}. Here, the device is tuned such that the left-dot ST splitting is much smaller than the right-dot ST splitting. Figure~\ref{fig:fig4}a shows the ST splittings extracted from pulsed-gate spectroscopy measurements of both dots, similarly to Fig.~\ref{fig:fig2}c. In this device tuning, the left-dot ST splitting increases from 34 $\mathrm{\mu eV}$ to 68 $\mathrm{\mu eV}$ along the P3 transition line and the right-dot ST splitting increases from 316 $\mathrm{\mu eV}$ to 418 $\mathrm{\mu eV}$ along the P5 transition line. The left-dot ST splittings are quite small compared to typical ST splittings in Ge/SiGe heterostructures, but not unprecedented. ST splittings as small as 20 $\mathrm{\mu eV}$ have been measured in DQDs in Ge/SiGe~\cite{De-Palma2024-nb} and even smaller ST splittings have been observed in DQDs in GaAs~\cite{Jang2021-rv}. These small ST splittings may be a result of Wigner molecule states in the DQD~\cite{Yannouleas2007-ql,Corrigan2021-er,Abadillo-Uriel2021-co,Ercan2021-qu,Yannouleas2022-xp,Yannouleas2022-sm,Yang2025-fw}.

The PAT and pulsed-gate spectroscopy data in this tuning are fit in the same way as the previous measurements, except the crosstalk is only included for the left-dot ST splitting. This is because the right-dot ST splitting is too large for the PAT measurements to be sensitive to its changes. Figure~\ref{fig:fig4}b compares the fitting results to experimental PAT measurements for selected microwave excitation frequencies. Here, curvature in the PAT data is visible in (2,1) as expected from the ST splittings. The contours from the global fit well replicate these features.

Figure~\ref{fig:fig4}c,d present a final device operating regime in which the ST splittings are tuned such that there is curvature in the PAT data in both the (2,1) and (1,2) charge configurations in the same microwave excitation frequency range. Figure~\ref{fig:fig4}c shows the ST splittings for both dots measured with pulsed-gate spectroscopy. In this tuning, the left-dot ST splitting increases from 143 $\mathrm{\mu eV}$ to 222 $\mathrm{\mu eV}$ along the P3 transition line and the right-dot ST splitting increases from 211 $\mathrm{\mu eV}$ to 370 $\mathrm{\mu eV}$ along the P5 transition line.

Figure~\ref{fig:fig4}d shows the experimental PAT data points and the contours from the fit at selected microwave excitation frequencies. Due to the more similar ST splittings of the two dots in this tuning, curvature is visible in both the (2,1) and (1,2) regions of the charge stability diagram. As such, the fitting procedure includes crosstalk for both ST splittings. As in the previous two cases, there is excellent agreement between the contours of best fit and the PAT data.

In summary, we study the gate-voltage dependence of the ST splittings of a DQD in a Ge/SiGe heterostructure. We find that making small changes to plunger gate voltages can result in large changes in the ST splittings, as shown by anomalous PAT resonances and pulsed-gate spectroscopy. We analyze both types of data together in a model that allows the ST splittings to vary as a linear function of the plunger gate voltages for the DQD, finding excellent agreement with the data. The results from the measurements and the global fits give insight into how the orbital energy levels in the DQD vary as their local electrostatic environment changes. This gate-voltage dependence is relevant for the readout and manipulation of hole spin qubits in Ge/SiGe heterostructures, especially qubits that can benefit from energy splitting tunability such as quantum dot hybrid qubits.

\textit{Notes}---The data that support the findings of this study are openly available in a \href{https://doi.org/10.5281/zenodo.19835276}{Zenodo repository}~\cite{Benson2026-zo}.

\textit{Acknowledgements}---We thank G. J. Bernhardt for experimental assistance. Research was sponsored in part by the Army Research Office (ARO) under Grant Nos. W911NF-23-1-0110 and W911NF-22-1-0257. ARH acknowledges support from the National Science Foundation (NSF) Graduate Research Fellowship Program under Grant No. 2137424 and the Graduate School and the Office of the Vice Chancellor for Research and Graduate Education at the University of Wisconsin-Madison with funding from the Wisconsin Alumni Research Foundation. The authors gratefully acknowledge the use of facilities and instrumentation in the Wisconsin Center for Nanoscale Technology. This Center is partially supported by the Wisconsin Materials Research Science and Engineering Center (NSF DMR-2309000) and by the University of Wisconsin–Madison. The views and conclusions expressed in this material are those of the authors and should not be interpreted as representing the official policies, either expressed or implied, of the ARO, the NSF, or the U.S. Government. The U.S. Government is authorized to reproduce and distribute reprints for Government purposes notwithstanding any copyright notation herein.

\textit{Competing Interests}---G.S. and M.V. are founding advisors of Groove Quantum BV and declare equity interests. The remaining authors declare no competing interests.
\vspace{1em}

\bibliography{main_bib.bib}

\clearpage

\section*{Supplemental material for `Large quantum dot energy level shifts in anomalous photon-assisted tunneling'}

\setcounter{figure}{0}

\subsection*{S1. PAT data supporting Fig.~\ref{fig:fig1}c}

\begin{figure}[ht!]
\renewcommand{\thefigure}{S\arabic{figure}}
\includegraphics[width=3.375in]{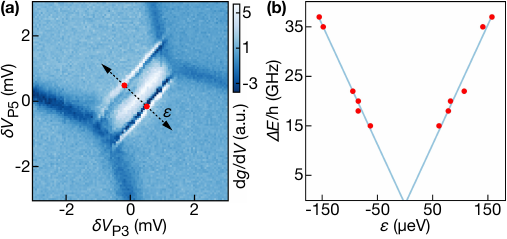}
\caption{\label{fig:figS1} PAT when the device is tuned such that the excited state anti-crossings are far from the ground-state anti-crossing in detuning. (a) A charge stability diagram showing photon-assisted tunneling from an applied 15 GHz microwave excitation. The dashed line indicates the detuning ($\varepsilon$) axis and the red dots show the position of the PAT resonances in detuning. (b) The position of the PAT resonances in detuning vs.\ the PAT resonance frequency. The data show a linear trend on either side of the polarization line, indicating that the excited state anticrossings are far from the polarization line in detuning.}
\end{figure}

The PAT data shown in Fig.~\ref{fig:fig1}c is acquired in a device tuning where the excited state anticrossings occur at large positive or negative detunings. As such, the PAT data in the frequency range we probe is not sensitive to the ST splittings or to the gate-voltage dependence of those splittings in this tuning. Figure~\ref{fig:figS1}a is the same data shown in Fig.~\ref{fig:fig1}c. The dashed line indicates the detuning axis, and the red dots denote the position of the straight PAT lines in this measurement. Figure~\ref{fig:figS1}b presents the position of the PAT resonances in detuning for multiple such stability diagrams measured while applying microwave tones with frequencies ranging from 15 GHz to 37 GHz. This data shows linear trends in both the negative and positive detuning directions. Therefore, we conclude that the excited state anticrossings are far from zero detuning, because if they were closer it would cause the trend in the PAT data to deviate from linear.

\subsection*{S2. Compensating for a device shift during the PAT measurements}

\begin{figure}[ht!]
\renewcommand{\thefigure}{S\arabic{figure}}
\includegraphics[width=3.375in]{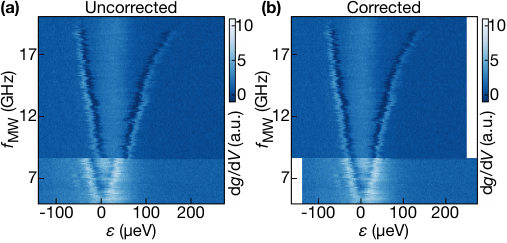}
\caption{\label{fig:figS2} Compensating for a shift in the device that occured while measuring the PAT data shown in Fig.~\ref{fig:fig3}c. (a) The uncorrected PAT data showing the device shift. (b) The same data after applying a constant offset in detuning to correct for the device shift.}
\end{figure}

While measuring the PAT data presented in Fig.~\ref{fig:fig3}c, there was a charge jump in the device that caused the PAT data to shift in detuning. This shift is clearly visible in the PAT measurement shown in Fig.~\ref{fig:figS2}a. We compensate for the shift by applying a constant detuning offset to the data taken after the charge jump occurred. As shown in Fig.~\ref{fig:figS2}b, this offset successfully corrects for the device shift.

We do not include the data shown in Fig.~\ref{fig:figS2} in the dataset presented in Fig.~\ref{fig:fig3}c due to the lower SNR after the device shift, but we apply the same offset to the following measurements in the dataset to account for the effect of the device shift on those measurements. 

\subsection*{S3. Extracting resonance frequencies from PAT data}

\begin{figure*}[ht!]
\centering
\renewcommand{\thefigure}{S\arabic{figure}}
\includegraphics[width=5.32in]{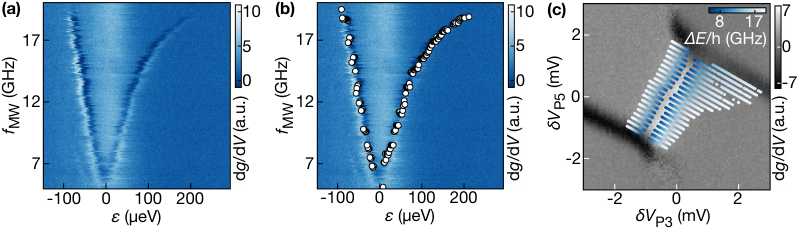}
\caption{\label{fig:figS3} Extracting resonance frequencies from PAT data. (a) A scan showing the PAT resonances of the DQD while sweeping the detuning and the frequency of the applied microwave excitation. (b) The same dataset with the extracted peak positions overlaid. (c) A map of the PAT resonance frequencies as a function of both plunger gate voltages compiled from data taken from the dataset in (a) and 19 additional measurements taken along offset lines in the stability diagram.}
\end{figure*}

To map out the dependence of the DQD PAT resonance frequencies on the plunger gate voltages, we combine information from several PAT measurements. Figure~\ref{fig:figS3}a shows one of these PAT measurements. Here, we pick a line in the stability diagram that is parallel with the detuning axis, and we measure the position of the PAT resonances along that line as we vary the frequency of the applied microwave excitation.

To extract the PAT resonances from the experimental data scans, we first process the data using a combination of morphological filtering, total variance denoising~\cite{Chambolle2004-lk}, and ridge filtering~\cite{Frangi1998-nd} implemented in scikit-image~\cite{van-der-Walt2014-qp}. We extract the peaks via a peak finding routine implemented using SciPy~\cite{Virtanen2020-oh}. Figure~\ref{fig:figS3}b shows the extracted peaks from the dataset in Fig.~\ref{fig:figS3}a overlaid on the data scan. We map out the PAT resonances over a two-dimensional area in the stability diagram by repeating the measurements and analysis along various lines in voltage space parallel to the detuning axis. The data is compiled in Fig.~\ref{fig:figS3}c, which shows data from 20 scans spanning the area between both ends of the polarization line.

\subsection*{S4. Data fitting}

To fit the points extracted from the PAT measurements, we use the following Hamiltonian as a model for the energy levels in the DQD~\cite{Shi2014-tl}:

\begin{equation}
    H=\begin{pmatrix}
        \varepsilon/2 & 0 & \Delta_1 & \Delta_2 \\
        0 & \varepsilon/2+E_L(V_{P3},V_{P5}) & \Delta_3 & \Delta_4 \\
        \Delta_1 & \Delta_3 & -\varepsilon/2 & 0 \\
        \Delta_2 & \Delta_4 & 0 & -\varepsilon/2 +E_R(V_{P3},V_{P5}) \\
    \end{pmatrix}
\end{equation}

Here, $\varepsilon$ is the double dot detuning, $\Delta_1$, $\Delta_2$, $\Delta_3$, $\Delta_4$ are the tunnel couplings, and $E_L$, $E_R$ are the ST splittings. The detuning can be calculated as follows:

\begin{equation}
    \varepsilon=\alpha_{P3}^\varepsilon(V_{P3}-V_{P3,0})-\alpha_{P5}^\varepsilon(V_{P5}-V_{P5,0}),
\end{equation}
where $\alpha_{P3}^\varepsilon$ ($\alpha_{P5}^\varepsilon$) is the lever arm describing the effect of the P3 (P5) gate voltage on the double dot detuning and $(V_{P3,0},V_{P5,0})$ describes a reference point on the polarization line.

We include the gate-voltage dependence of the ST splittings using the following equations:

\begin{multline}
    E_L(V_{P3},V_{P5}) = \alpha_{P3}^{R} \Big(\frac{1}{1+\beta_{P5}^L}\Big)\Big(\frac{\delta \Delta V_{L}}{\delta V_{P3}}(V_{P3}-V_{P3,1})\\+\Delta V_{L,1}+\beta_{P5}^L\Big(\frac{\delta \Delta V_{L}}{\delta V_{P5}}(V_{P5}-V_{P5,2})+\Delta V_{L,2}\Big)\Big)
\end{multline}

\begin{multline}
    E_R(V_{P3},V_{P5}) = \alpha_{P5}^{L} \Big(\frac{1}{1+\beta_{P3}^{R}}\Big)\Big(\frac{\delta \Delta V_{R}}{\delta V_{P5}}(V_{P5}-V_{P5,1})\\+\Delta V_{R,1}+\beta_{P3}^R\Big(\frac{\delta \Delta V_{R}}{\delta V_{P3}}(V_{P3}-V_{P3,2})+\Delta V_{R,2}\Big)\Big),
\end{multline}
where $\delta \Delta V_{L}/\delta V_{P3}$ ($\delta \Delta V_{R}/\delta V_{P5}$) and $\Delta V_{L,1}$ ($\Delta V_{R,1}$) are parameters from the linear fits to pulsed-gate spectroscopy results measured along the left (right) dot transition lines with respect to the P3 (P5) gate voltages, such as the linear fits shown in Fig.~\ref{fig:fig2}c. To include the effects of crosstalk on the left (right) dot ST splitting from the P5 (P3) plunger gate voltage, we also include $\delta \Delta V_{L}/\delta V_{P5}$ ($\delta \Delta V_{R}/\delta V_{P3}$) and $\Delta V_{L,2}$ ($\Delta V_{R,2}$). These are also determined by linear fits to the pulsed-gate spectroscopy results, except here the fitting is done with respect to the opposite plunger gate voltage. The parameters $\beta_{P5}^L$ ($\beta_{P3}^R$) quantify the size of the crosstalk effect of the P5 (P3) plunger gate voltage on the left (right) dot ST splitting, and they are fitting parameters.

The lever arm $\alpha_{P3}^L$ ($\alpha_{P5}^R$) describes the effect of the P3 (P5) gate voltage on the chemical potential of the left (right) dot in the double dot. These lever arms are calculated from the detuning lever arms and slopes of transition lines in the stability diagram as follows:

\begin{equation}
    \alpha_{P5}^R = \frac{\alpha_{P5}^\varepsilon+\Big\vert\Big(\frac{\delta V_{P3}}{\delta V_{P5}}\Big)_L\Big\vert\alpha_{P3}^\varepsilon}{1-\Big\vert\Big(\frac{\delta V_{P3}}{\delta V_{P5}}\Big)_L\Big\vert \Big\vert\Big(\frac{\delta V_{P5}}{\delta V_{P3}}\Big)_R\Big\vert}
\end{equation}

\begin{equation}
    \alpha_{P3}^L=\alpha_{P3}^\varepsilon+\Big\vert\Big(\frac{\delta V_{P5}}{\delta V_{P3}}\Big)_R\Big\vert\alpha_{P5}^R
\end{equation}
where $\vert (\delta V_{P3}/\delta V_{P5})_L\vert$ ($\vert (\delta V_{P5}/\delta V_{P3})_R\vert$) is the slope of the left (right) dot transition line in the stability diagram.

\begin{table}
\centering
\begin{tabular}{|l|l|l|l|} 
\hline
   & Figure 3 & Figure 4b & Figure 4d  \\ 
\hline
$\Delta_1$ (GHz) & 2.40    & 3.63     & $0^*$      \\ 
\hline
$\Delta_2$ (GHz) & 5.01    & 3.55     & 13.44      \\ 
\hline
$\Delta_3$ (GHz) & 7.28   & 3.99     & 8.65      \\ 
\hline
$\Delta_4$ (GHz) & 17.62   & $0^*$     & 32.66      \\ 
\hline
$\varepsilon_0$ ($\mu\mathrm{eV}$) & 20.10    & $0^*$     & 5.61      \\ 
\hline
$\alpha_{P3}^\varepsilon$ (eV/V) & 0.099    & 0.035     & 0.119     \\ 
\hline
$\alpha_{P5}^\varepsilon$ (eV/V) & 0.060    & 0.058     & 0.091      \\ 
\hline
$\beta_{P5}^L$ & $0^*$    & 0.565    & -0.312      \\ 
\hline
$\beta_{P3}^R$ & -0.082    & $0^*$     & 0.605      \\
\hline
\end{tabular}
\caption{\label{tab:tab1} Parameter values from PAT data fitting.}
\end{table}

When performing the curve fitting, we fit all the PAT data in the two-dimensional map simultaneously. To do this, we use the Levenberg–Marquardt algorithm in LMFIT~\cite{Newville2025-jo}. Notably, since the microwave excitations applied to the device provide an absolute energy scale through their photon energy, the PAT measurements are self-calibrating, and we determine the detuning lever arms through the fitting procedure.

The parameter values from the fitting for the three different device tunings are summarized in Table~\ref{tab:tab1}. The parameter $\varepsilon_0$ is an offset for the detuning. Additionally, when a parameter value is marked with $0^*$ this means that it is constrained to zero for the fitting in that case. This constraint is enforced for the crosstalk parameters $\beta_{P5}^L$ ($\beta_{P3}^R$) in cases where we include crosstalk for only one of the ST splittings in the model, and for the tunnel couplings when it is necessary to properly constrain the fit.

\subsection{S5. Measuring the hole temperature}

\begin{figure}[ht!]
\renewcommand{\thefigure}{S\arabic{figure}}
\includegraphics[width=3.375in]{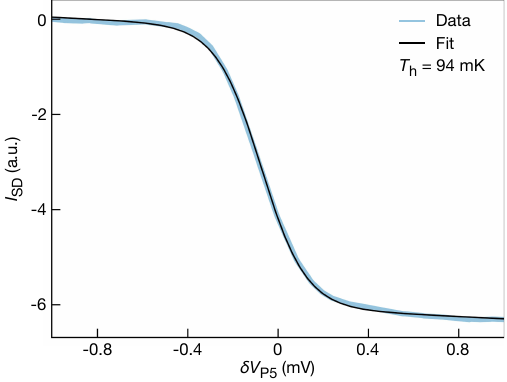}
\caption{\label{fig:figS4} Hole temperature measurement. The lineshape of a reservoir transition in the P5 dot is measured and fit with a Fermi-Dirac distribution. The hole temperature from this fit is 94 mK.}
\end{figure}

To determine the hole temperature, we measure a reservoir transition in the P5 dot when the device is tuned such that the transition line is not broadened by the tunnel coupling between the P5 dot and the reservoir. We subtract the background slope from the data such that that slope becomes zero before and after the transition line, and then we numerically integrate it to give a quantity that is analogous to the charge sensor current, which is plotted in Fig.~\ref{fig:figS4}. We fit the data with a Fermi-Dirac distribution, and from this fit we extract a hole temperature of 94 mK.

\subsection*{S6. Experimental setup}

\begin{figure}[ht!]
\renewcommand{\thefigure}{S\arabic{figure}}
\includegraphics[width=3.375in]{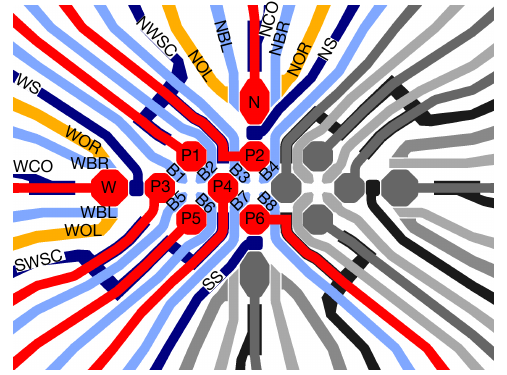}
\caption{\label{fig:figS5} A diagram of the device that is measured in this work with gate labels. The unlabeled gates are not wirebonded to the sample PCB and are left floating.}
\end{figure}

Figure~\ref{fig:figS5} is a diagram of the gate layout of the device measured here with all the relevant gates labeled. The device is a two-dimensional quantum dot array with ten dots and four charge sensors. The gate stack is nominally identical to the device measured in Ref.~\cite{John2025-ml}, but the Ge/SiGe heterostructure is grown on a silicon wafer as in Ref.~\cite{Lodari2021-sb}. The gates that are colored in grayscale and are left unlabeled are not wirebonded to the sample PCB and are instead left floating.

\begin{figure*}[ht!]
\centering
\renewcommand{\thefigure}{S\arabic{figure}}
\includegraphics[width=6.23in]{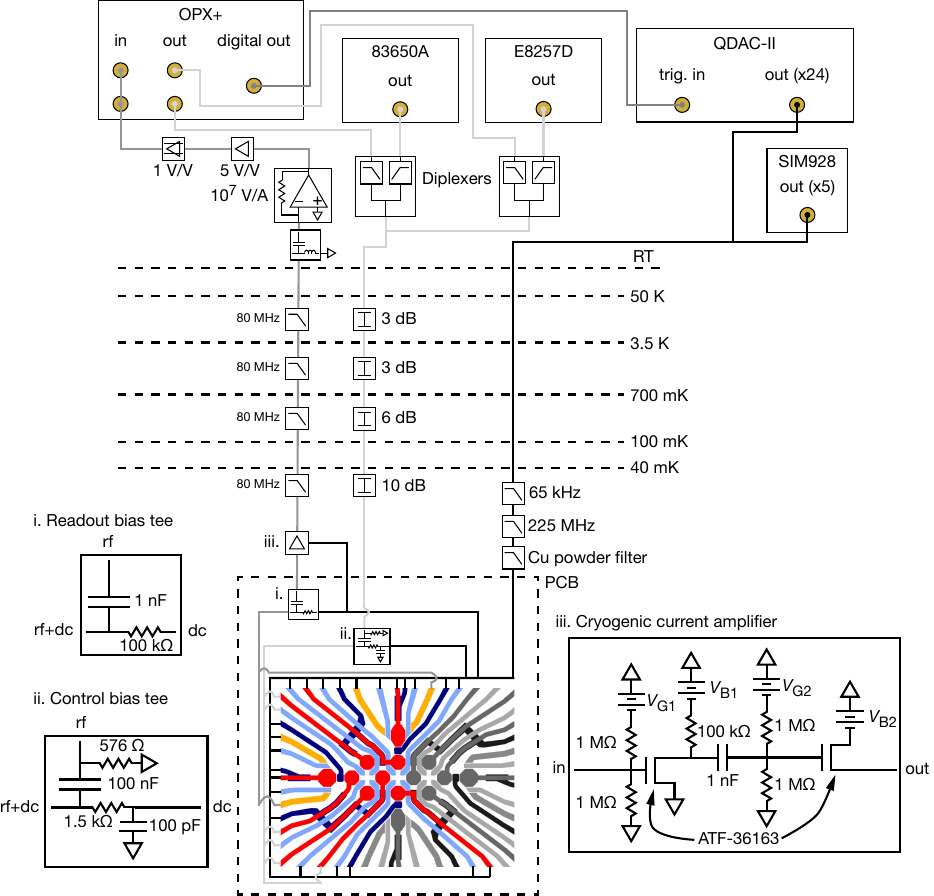}
\caption{\label{fig:figS6} A diagram of the experimental setup including the dilution refrigerator wiring and room temperature electronics. The insets show circuit diagrams for the on-board bias tees on the sample PCB and the cryogenic current amplifier used for readout.}
\end{figure*}

Figure~\ref{fig:figS6} shows the experimental setup, including the wiring of the dilution refrigerator and the room temperature instruments used in the measurements. The quantum dot device is mounted to the mixing chamber of a dilution refrigerator, and it is wirebonded to a printed circuit board (PCB). Static dc voltages, generated at room temperature by a QM QDAC-II, are supplied to the device through twisted-pair loom that is clamped to each plate in the dilution refrigerator before passing through RC+RF filters (QM QFilter-II) and Cu powder filters that are all thermalized to the mixing chamber plate. Microwave signals used for photon-assisted tunneling are generated by two microwave signal generators (HP 83650A and Agilent E8257D) at room temperature. These signals are combined with low-frequency signals from a QM OPX+ including lock-in modulations and square-wave pulses using diplexers (Marki Microwave MDPX-0305). These signals are transmitted to the device using coaxial cables in the dilution refrigerator that are thermalized at various temperature stages using attenuators. They are combined with dc signals on the PCB using bias tees, as shown in inset ii. Importantly, the rf line connected to the P3 gate is discontinuous somewhere in the dilution refrigerator. To apply high-frequency signals to the P3 dot, we instead use the rf line connected to the B1 barrier gate. This is possible because there is no dot tuned up under P1, so the signals we apply to the B1 gate primarily affect the P3 dot. 

For readout, the currents from ohmic contacts adjacent to the charge sensors are amplified and measured. The first amplification stage is a cryogenic current amplifier that is mounted to the mixing chamber of the dilution refrigerator in close proximity to the device. The readout signal is routed out of the device using a readout bias tee (inset i.) and connected to the cryogenic amplifier using an ultra-miniature stainless steel coaxial cable to minimize the thermal load on the device. The cryogenic current amplifier, shown in inset iii., is a two-stage amplifier that uses Avago ATF-36163 high-electron-mobility transistors (HEMTs). The amplifier is controlled by two bias voltages and two gate voltages, which are provided by four SRS SIM928 isolated voltage sources. The readout signal passes through low-pass filters at the same temperature stages as the attenuators (Mini-Circuits VLF-80+). The dc part of this current is shunted to ground using a bias tee at room temperature (PSPL 5546). The remaining current is then amplified by a Femto DHPCA-100 transimpedance amplifier and a SRS SR560 voltage amplifier. Finally, it passes through a fully differential amplifier (Analog Devices ADA-4945, powered by a SRS SIM928), which limits the signal to $\pm 0.5\,\mathrm{V}$. The differential output is routed into two analog inputs on the QM OPX+, which performs differential measurements on the two inputs.

\end{document}